\def\BEQ{\begin{eqnarray}}
\def\EEQ{\end{eqnarray}}
\def\BML{\begin{mathletters}}
\def\EML{\end{mathletters}}
\def\BE{\begin{equation}}
\def\EE{\end{equation}}
\def\NN{\nonumber}
\def\Sv{{\bf S}}
\def\e{\,{\rm e}}

\def\d{\partial}

\def\dr{{\rm d}}
\def\r{\right}
\def\l{\left}
\def\R{\rangle}
\def\L{\langle}
\def\p{\varphi}
\def\hf{{1\over2}}
\def\Tr{{\rm Tr}\,}
\def\ZZ{{\Bbb Z}}
\def\hf{{1\over2}}
\def\la{\lambda}
\def\La{\Lambda}
\def\t{\theta}
\def\Lag{{\cal L}}
\def\Jperp{J_\bot}

\def\phib{\skew5\bar\phi}
\def\Phib{\bar\Phi}
\def\psib{\skew5\bar\psi}
\def\zetab{\bar\zeta}
\def\Jb{{\bar J}}
\def\qb{{\bar q}}
\def\pb{{\bar p}}
\def\wb{{\bar w}}
\def\zb{{\bar z}}
\def\ab{{\bar a}}
\def\s{\sigma}
\def\bra#1{\L #1 |}
\def\ket#1{|#1\R}

\documentstyle[prb,aps,floats,amsfonts]{revtex}
\input epsf.tex
\begin{document}
\bibliographystyle{prsty}
\draft

\title{The spin-1 ladder : A bosonization study}

\author{D. Allen\cite{AllenAddress} and D. S\'en\'echal}

\address{Centre de Recherche en Physique du Solide et D\'epartement de
Physique,}
\address{Universit\'e de Sherbrooke, Sherbrooke, Qu\'ebec,
Canada J1K 2R1.}
\date{August 1999, {\tt CRPS-99-08}}

\maketitle
\begin{abstract}
We construct a field-theoretic description of two coupled spin-1
Heisenberg chains, starting with the known representation of a single spin-1
chain in terms of  Majorana fermions (or Ising models). After reexamining the
bosonization rules for two Ising models, taking particular care of order and
disorder operators, we obtain a bosonic description of the spin-1 ladder. From
renormalization-group and mean-field arguments, we conclude that, for a
small interchain coupling, the spin-1 ladder is approximately described
by three decoupled, two-frequency sine-Gordon models. We then predict
that, starting with decoupled chains, the spin gap decreases linearly
with interchain coupling, both in the ferromagnetic and antiferromagnetic
directions. Finally, we discuss the possibility of an incommensurate phase in
the spin-1 zigzag chain.
\end{abstract}
\section{Introduction}

Among the properties of spin ladders, the best known is the
reduction of order as we go from a single spin-$\hf$ Heisenberg chain to two
coupled chains~: the single spin-$\hf$ chain is critical (its
correlation length is infinite) whereas the spin-$\hf$ ladder has finite-range
correlations and an excitation gap, growing linearly with interchain
coupling $\Jperp$, at least for small $\Jperp$ (for a
review and further references, see Ref.~\onlinecite{Dagotto96}). This may seem
paradoxical because one would naively expect that coupling two quasi-ordered
chains would only increase the tendency to order, but a critical system like
the spin-$\hf$ Heisenberg chain is easily sent off-criticality by a
perturbation such as ladder coupling $\Jperp$. In this paper we will study the
corresponding spin-1 ladder (two coupled spin-1 chains), which is already
disordered and has a finite gap at $\Jperp=0$. On the contrary, we will argue
that the spin gap decreases as $\Jperp$ increases from zero, and does so for
both antiferromagnetic and ferromagnetic interchain couplings, thus giving the
gap $\Delta(\Jperp)$ a nonanalytic behavior (a cusp) at zero (cf.
Fig.~\ref{gapFIG} below). We will arrive at this conclusion after obtaining a
field-theoretic description of the spin-1 ladder in terms of six quantum Ising
models, or alternately in terms of three boson fields. The motivation for using
bosonization is that it offers a safer description of the system at weak
$\Jperp$, valid for both positive and negative $\Jperp$, and allows at the
same time for a description of the spin-1 zigzag chain, in which frustration
plays a r\^ole. Thus, at small $\Jperp$, this method is more general and
reliable than a sigma model description.  For a small antiferromagnetic
interchain coupling, the drop in the gap as a function of $\Jperp$ was
already noticed in Monte Carlo simulations and accounted for with a nonlinear
sigma model description of the spin-1 ladder.\cite{Senechal95} 

We will consider the spin-1 ladder as a perturbed critical model, so that the
low-energy description of the system will be a perturbed conformal field
theory. The critical model used as a starting point is a pair of
decoupled biquadratic spin chains, with Hamiltonian
\BE\label{critspin}
H_0=\sum_{\alpha,i}\l\{ \Sv_{\alpha,i}\cdot\Sv_{\alpha,i+1}
-\l( \Sv_{\alpha,i}\cdot\Sv_{\alpha,i+1}\r)^2 \r\}
\EE
where $\Sv_{\alpha,i}$ is a spin-1 operator at site $i$ on chain
$\alpha$ ($\alpha=1,2$). At this critical point the two chains are
decoupled, each chain being described by an integrable
model\cite{Babujian82,Kennedy92} which is equivalent in the continuum
limit to a level-2 su(2) Wess-Zumino-Witten (WZW) model.\cite{Affleck87} 
We then need to consider the following perturbation:
\BE\label{perturb}
H_I=(1+\eta)\sum_{i,\alpha}\l(\Sv_{\alpha,i}\cdot\Sv_{\alpha,i+1}\r)^2
+\hf\Jperp\sum_i
\Sv_{1,i}\cdot\l[(1+\delta)\Sv_{2,i}+(1-\delta)\Sv_{2,i+1}\r]
\EE
When $(1+\eta)>0$, the first term brings us back to the Heisenberg point
($\eta=0$). The interchain interaction, of strength $\Jperp$, is that of
a ladder ($\delta=\pm 1$) or of a zigzag chain ($\delta=0$). We will
proceed by (i) constructing a continuum description of the interaction
in term of WZW models and (ii) finding out the behavior of this
perturbed WZW model by field-theoretic methods, mainly through
representations in terms of Ising models (fermionization) and sine-Gordon
models (bosonization).

\begin{figure}
\epsfxsize 8cm\centerline{\epsfbox{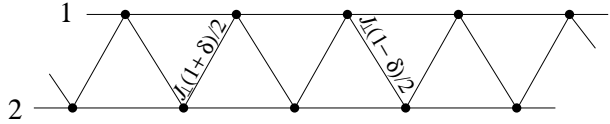}}
\caption{Schematic illustration of the coupled spin chains with the
various couplings, normalized to the intra-chain coupling.}
\end{figure}

This paper is organized as follows. In Sect.~\ref{S-1chain}, we review the
field-theoretic description of a single spin-1 chain, in particular its
representation in terms of three Majorana fermions (or Ising models). In
Sect.~\ref{S-2chains}, we write down a representation of two coupled spin-1
chains in terms of three bosons, using the bosonization formulas for pairs of
Ising models given in Appendix~\ref{appIsing}. In Sect.~\ref{S-gap}, the
behavior of the spin gap as a function of interchain coupling is inferred from
this bosonized description. In Sect.~\ref{S-zigzag}, the spin-1 zigzag chain is
considered instead, and a weak interchain coupling is argued to cause a
short-distance incommensurability, i.e., a displacement of the minimum of the
one-magnon spectrum from $q=\pi$.

\section{Continuum description of the single spin-1 chain}
\label{S-1chain}

\subsection{Phase diagram}
\label{phasediagS}

Let us first review the phase diagram of the biquadratic spin-1
chain:\cite{Fath93}
\BE\label{biquad}
H=\sum_i\l\{\Sv_i\cdot\Sv_{i+1}
+\eta\l(\Sv_i\cdot\Sv_{i+1}\r)^2\r\}
\EE
At $\eta=-1$, the Hamiltonian is integrable and has gapless modes at $k=0$
and $k=\pi$. It is also integrable at $\eta=1$ and has then gapless modes at
$k=0$ and $k=\pm2\pi/3$. If $\eta<-1$ we have a dimerized phase characterized
by two degenerate ground states with a finite gap. On the other hand, in the
interval $\eta\in~(-1,1)$ the spectrum has a singlet ground state with a
finite gap. This is the so-called Haldane phase, characterized by the
spontaneous breakdown of a
$\ZZ_2\times\ZZ_2$ symmetry.\cite{Nijs89,Kennedy92B} This
breakdown implies a four-fold degenerate ground state in an open chain, but
these different ground states differ only by the spins at the
ends of the chain, and in this sense they are equivalent in the thermodynamic
limit. The excitations are solitons switching from one ground state
at $x\to-\infty$ to another ground state at $x\to\infty$. Related to the
symmetry breaking is a dilute antiferromagnetic order; schematically:
\BE
+\, 0\cdots 0\, -\, 0\cdots 0\, +\, 0\cdots 0\, -
\EE
This order is defined by an alternation of sites with $S_z=1$ and
$S_z=-1$, with some $S_z=0$ sites in between. It can be measured by the
so-called string order parameter:\cite{Nijs89}
\BE\label{string}
{\cal O}^z=\lim_{m-n\to\infty}\l\L S_n^z\exp
\l(i\pi\sum_{k=n+1}^{m-1} S_k^z\r) S_m^z\r\R
\EE
This order parameter and the gap are maximal at $\eta=1/3$,
where the VBS-like ground state is exactly known.\cite{Affleck87B} The gap
grows monotonically from $\eta=-1$ to $\eta=0$ without phase transition, and
thus we may consider the Heisenberg point ($\eta=0$) as a perturbation of the
critical point ($\eta=-1$). 

Note also that incommensurability develops starting at $\eta\approx 0.4$: the peak
in the spin-spin correlation function moving from $k=\pi$ to $k=2\pi/3$ at
$\eta=1$.\cite{Golinelli98} This last transition point is described by an SU(3)
generalization of the Kosterlitz-Thouless phase transition.\cite{Itoi97}

\subsection{Field-theoretic description}

The critical point ($\eta=-1$) is equivalent, in the low-energy limit, to 
a conformal field theory: the
su(2) Wess-Zumino-Witten (WZW) model at level $k=2$, plus a marginally
irrelevant perturbation.\cite{Affleck87} This WZW model contains two scaling
fields~: a spin doublet $g_{mn}$ ($m,n\in\{-\hf,\hf\}$) with left and
right conformal dimensions $(\frac3{16},\frac3{16})$ and a spin triplet
$\Phi_{mn}$ ($m,n\in\{-1,0,1\}$) with dimensions $(\frac12,\frac12)$. They are
respectively $2\times2$ and
$3\times3$ matrix fields. The link between the spin chain and the WZW model is
given by the following representation of the spin operators in the continuum
limit:\cite{Affleck87,Cabra98}
\BE\label{spinWZW}
{1\over a_0}S^a(x)={1\over 2}\l(
J^a(x)+{\Jb}^a(x)\r) +(-1)^{x/a_0}\Theta g^a(x)
\EE
where $a_0$ is the lattice constant, $\Theta$ a nonuniversal constant, $J^a$
and $\Jb^a$ are the right and left su(2) currents and $g^a$ is defined in
terms of Pauli matrices as
\BE
g^a={1\over\sqrt 2}\Tr(\s^a g) = {1\over\sqrt 2}\sum_{m,n}\s^a_{mn} g_{nm}
\EE
The currents ($J^a$, $\Jb^a$) and the field $g^a$ correspond to the soft modes
of the spin chain near $k=0$ and $\pi$, respectively. 

For $1+\eta$ not too large, the spin chain may be described by the
above WZW model, plus the following perturbation:\cite{Affleck87}
\BE\label{perturbWZW}
\Lag_1 = m\Tr\Phi-\la_1 J^a\Jb^a
\EE
where a summation over repeated indices is implicit and $-m$ is proportional to
$(1+\eta)$ ($m$ is negative in the Haldane phase). The second term is the
marginally irrelevant perturbation alluded to above (if $\la_1>0$). On the
other hand, the first term ($\Tr\Phi$) is relevant, with scaling dimension 1,
and leads to a gap proportional to $|m|\propto |1+\eta|$.

There is an interesting equivalence between the $k=2$ su(2) WZW model and
three quantum Ising models,\cite{Zamolodchikov86} and so we will not have to
deal with the WZW model directly. This equivalence is defined by the following
relations:
\BE\label{WZWisingd}
J^a={-i\over\sqrt 2}\epsilon_{abc}\psi_b\psi_c\hskip
2truecm {\Jb}{}^a={-i\over\sqrt 2}\epsilon_{abc}\psib_b\psib_c
\EE
\BE
\Phi_1={\zeta\over\sqrt 2}\l(-\psi_1+i\psi_2\r)\hskip 1.5truecm
\Phi_0=\zeta\psi_3\hskip 1.5truecm
\Phi_{-1}={\zeta\over\sqrt 2}\l(\psi_1+i\psi_2\r)
\EE
\BE
{\bar\Phi}_1={\zetab\over\sqrt
2}\l(-\psib_1-i\psib_2\r)\hskip 1.5truecm
{\bar\Phi}_0=\zetab\psib_3\hskip 1.5truecm
{\bar\Phi}_{-1}={\zetab\over\sqrt 2}\l(\psib_1-i\psib_2\r)
\EE
\BE\label{WZWisinge}
g^0=\sqrt 2\s_1\s_2\s_3\hskip 2truecm g^a=-\sqrt
2\s_a\mu_{a+1}\mu_{a+2}
\EE
where the latin index goes from 1 to 3; $\psi_a$ and $\psib_a$ are
respectively the right and left fermions associated with each Ising model (see
Appendix~\ref{appIsing}). $\s_a$ and $\mu_a$ are the order and disorder fields
of each Ising model. The $3\times 3$ matrix field $\Phi_{nm}$ is here
factorized as $\Phi_{nm}\equiv\Phi_n\Phib_m$. The constants $\zeta$ and
$\zetab$ are such that their product is $\zeta\zetab=i$. Note that our
relations differ slightly from those given by Fateev and
Zamolodchikov.\cite{Zamolodchikov86}  The action of the WZW model in imaginary
time becomes simply that of free Majorana fermions:
\BE
S_{\rm WZW}={1\over 2\pi}\int \dr x\dr\tau\; (
\psi_a{\bar\d}\psi_a+\psib_a\d\psib_a )
\EE
where $\d=(\d_\tau-i\d_x)/2$ and ${\bar\d}=(\d_\tau +i\d_x)/2$ (in order to
lighten the notation, the characteristic velocity $v$ of the WZW model has
been set to unity). The perturbation (\ref{perturbWZW}) becomes:
\BE
\Lag_1=m\psi_a\psib_a-\la_1\psi_a\psib_a\psi_b\psib_b
\EE
Except for the marginally irrelevant term, the spin chain is thus equivalent to
three Majorana fermions of mass $m$. This description of the spin-1
chain has been used to study the effect of a magnetic field on the low energy
spectrum.\cite{Tsvelik90} 

The representation (\ref{WZWisingd}--\ref{WZWisinge}) of the WZW fields is
invariant under the following changes (for $a=1,2,3$ simultaneously):
\BE\label{gaugesym}
\psi_a\to -\psi_a\qquad
\psib_a\to -\psib_a\qquad
\mu_a\to -\mu_a\qquad
\s_a\to \s_a
\EE
This is related to the absence of fermionic field in the WZW model. This
`gauge' symmetry accounts for the expected degeneracy of the ground state
near the critical point in open chains. Specifically, recall that $m<0$ in
the Haldane phase. In our formulation,
this corresponds to the disordered phase of the Ising models (see
Appendix~\ref{appIsing}) and the expectation value of the disordered
operators is nonzero: $\L\mu_a\R\ne0$. In this phase each Ising model has a
doubly-degenerate ground state, associated to different spin configurations
at the ends of the open Ising chain. The two ground states differ in the sign
of $\L\mu_a\R$. For the spin chain, this degeneracy implies an apparent
eight-fold ($8=2^3$) degeneracy, but the gauge invariance (\ref{gaugesym})
reduces this to a physical four-fold degeneracy. These different ground
states come from the breakdown of the hidden
$\ZZ_2\times\ZZ_2$ nonlocal symmetry alluded to above, and are physically
equivalent in the thermodynamic limit. In this Ising model description of the
spin chain, the elementary excitations are kinks switching from one value of
$\L\mu_a\R$ at
$x\to-\infty$ to its opposite at
$x\to\infty$. On the other hand, F\'ath and S\'olyom\cite{Fath93} have shown
that the excitations of the Heisenberg model are solitons connecting the
ground states with different values of the string parameter (\ref{string}). We
are thus led to identify these solitons with the kinks of the Ising model.

We can do the same exercise for $m>0$ (or $\eta<-1$). We are now in the ordered
phase of the Ising models: $\L\s_a\R\neq 0$. Such an expectation value is
already invariant under the gauge change (\ref{gaugesym}) and therefore there
are really 8 physically different ground states for the open chain. A hidden
$\ZZ_2\times\ZZ_2$ symmetry breaking is again expected and so these 8
different ground states will be locally equivalent to two distinct ground
states in the thermodynamic limit, corresponding to the expected dimerized
state.

\section{Bosonization}
\label{S-2chains}

Using the continuum description (\ref{spinWZW}) of the spin operators, 
we obtain the following Lagrangian density from the Hamiltonian
(\ref{critspin},\ref{perturb}), in terms of WZW fields:
\BEQ\label{lagcoupled}
\Lag &=& \Lag_{\rm WZW}[g]+\Lag_{\rm WZW}[g']
+m\Tr\Phi+m\Tr\Phi'
-\la_1(J^a\Jb^a+J'^a\Jb'^a)\NN\\
&& +\la_2(J^a J'^a+\Jb^a\Jb'^a) +\la_3(J^a\Jb'^a + \Jb^a J'^a) 
+\la g^ag'^a +\rho
\l(g^a\d_x g'^a-(\d_xg^a)g'^a\r) 
\EEQ
The unprimed fields correspond to the first chain and the primed fields to
second chain. The first three terms represent the intrachain interaction and
the last four terms the interchain coupling. The interchain couplings
$\la_2$, $\la_3$, $\la$ and $\rho$ are respectively proportional to
$\Jperp$, $\Jperp$, $\Jperp\delta$ and $\Jperp(1-\delta^2)$ at high-energy,
but they renormalize differently towards low-energy. The last term has been
omitted in previous work on the zigzag spin chain,\cite{White96,Allen97} but
since it respects the symmetry of the microscopic model, it is not forbidden
and we expect $\rho$ to be nonzero. Note that $\rho$ must be zero for the spin
ladder, whereas $\la$ vanishes for the pure zigzag chain. In the following we
will consider the Haldane phase only so that $m$ is negative.

The Lagrangian (\ref{lagcoupled}) is difficult to study in terms of WZW
fields. The simplest information we may extract from it is the scaling
dimension of the various perturbations, from those of the various WZW fields.
Thus, the interchain couplings $\la_2$, $\la_3$, $\la$ and $\rho$ respectively
have scaling dimension 2, 2, $\frac34$, and $\frac74$. Moreover, the couplings
$\la_2$ and $\rho$ have conformal spin. By itself, a relevant coupling $g$ of
scaling dimension $\gamma<2$ and zero conformal spin is expected to produce a
gap of order $\Delta\sim g^{1/(2-\gamma)}$. Thus, at the in-chain critical
point ($\eta=-1$), the interchain coupling $\la$ would open a gap of
order
\BEQ
\Delta(\la) \sim \la^{4/5}
\EEQ
in the spin-1 ladder.

However, far from the critical point, the
WZW model is of little help in predicting the behavior of the gap and
the fermionic language seems more appropriate. Using the
representation (\ref{WZWisingd}--\ref{WZWisinge}), we can express the
Lagrangian density (\ref{lagcoupled}) in terms of Majorana fermions, order and
disorder fields. Unfortunately, the resulting expression is not easy to
study since it contains a mixture of fields that are mutually nonlocal
(the order and disorder operators).

An interesting way to deal with the Lagrangian
(\ref{lagcoupled}) is bosonization. The 2D Ising models may be
bosonized by pairing them (see Appendix~\ref{appIsing}). The natural way to 
bosonize the ladder is to pair an Ising model describing one chain with its
twin on the other chain. Using the relations
(\ref{WZWisingd}--\ref{WZWisinge}), (\ref{lagcoupled}) and (\ref{bosonising}),
we obtain the following Lagrangian density for two coupled spin-1 chains:
\BEQ\label{Lagboson}
\Lag &=& \Lag_0 + \Lag_1 + \Lag_2 + \Lag_3 + \Lag_\rho + \Lag_\la \NN\\
\Lag_0 &=& \sum_{a=1,2,3}
\Big[ {1\over 8\pi}\l[ (\d_\tau\p_a)^2+(\d_x\p_a)^2\r] 
-2m\cos{\p_a}\Big]\NN\\
\Lag_1 &=&  16\la_1\sum_{a=1,2,3}\big(\cos\p_{a+1}\cos\p_{a+2}
+\cos\t_{a+1}\cos\t_{a+2}\big)\NN\\
\Lag_2 &=& 4\la_2\sum_{a=1,2,3}
\big(\d_\tau\p_{a+1}\d_\tau\p_{a+2}-\d_x\p_{a+1}\d_x\p_{a+2}\big)\NN\\
\Lag_3 &=& -8\la_3\sum_{a=1,2,3}\big(\sin\p_{a+1}\sin\p_{a+2}+
\sin\t_{a+1}\sin\t_{a+2}\big)\NN\\
\Lag_\la &=& 4\sqrt
2\la\sum_{a=1,2,3}\cos{\p_a\over2}\sin{\p_{a+1}\over2}
\sin{\p_{a+2}\over2} \NN\\
\Lag_\rho &=& -4\sqrt 2 \rho \sum_{a=1,2,3}\cos\theta_a\l[
\cos{\p_a\over 2}\sin{\p_{a+1}\over 2}\sin{\p_{a+2}\over 2}
-\sin{\p_{a}\over 2}
\sin\l({\p_{a+1}+\p_{a+2}\over 2}\r)\r]
\EEQ
where $\t_a$ is the boson dual to $\p_a$. To shorten the  expression, we
have adopted a periodic condition on the index $a$, i.e., $a+3\equiv a$. The
twist term $\Lag_\rho$, the trickiest to bosonized, has been inferred from
the representation (\ref{EMtens}) of the stress-energy tensor for each Ising
model, plus the usual operator product expansion (OPE) between the
energy-momentum tensor and a conformal field.

Thus, we have transformed the problem into a system of three perturbed
sine-Gordon models, although the simultaneous presence of the
bosons $\p_a$ and of their dual fields $\t_a$ makes some perturbations
nonlocal. However, as we will see, the most relevant perturbation is local
and makes the problem tractable in this language. Note that our
normalization is such that $\cos(\beta\p_a)$ is marginal for $\beta=\sqrt 2$,
and thus bound states appear in the sine-Gordon model for $\beta<1$. Also, for
$\beta<\sqrt2$, the $\p_a\to\p_a+2\pi$ symmetry is spontaneously broken and we
have to consider fluctuations around one of the minima of the potential. 
However, it is important to keep in mind that our bosonization procedure is
from the start invariant under the translation $\p_a\to\p_a+4\pi$, and this
$4\pi$-periodicity property must be regarded as a constitutive constraint
imposed on the sine-Gordon models. Thus, each sine-Gordon model in $\Lag_0$
has {\it two} inequivalent ground states, associated with the minima
$\p_a=\pm\pi$ of the potential (for $m<0$). The spontaneous breakdown of the symmetry $\p_a\to
\p_a+2\pi$ implies a nonzero expectation value for the operators
$\sin\p_a/2$ (the disorder operators $\mu_a$) and $\cos\p_a$. Moreover,
this breakdown becomes explicit if the perturbation $\Lag_\la$ or
$\Lag_\rho$ is added. This symmetry breaking of the three sine-Gordon models
corresponds in fact to the hidden symmetry breaking in the spin-1 chain (cf.
Sect.~\ref{phasediagS}). Therefore the different choices of the
ground state ($\p_a=\pm\pi$) are equivalent, since the different ground
states of the spin-1 chain are equivalent in the thermodynamic limit.
Finally, let us recall that the elementary excitations of each sine-Gordon
model have finite mass and correspond to the kink and antikink connecting
the two different ground states. The charge conjugation changing kink into
antikink corresponds to the following transformation:
\BE\label{cconjug}
\p(x)\to 2\pi-\p(x)~{\rm mod}~4\pi
\EE

The presence of nonzero expectation values for the operators $\sin\p_a/2$
and $\cos\p_a$ implies that more relevant terms may be generated from the
perturbations (\ref{Lagboson}). If we call $\alpha$ the expectation value of
$\sin\p_a/2$, say in the $\p=\pi$ ground state, and $\alpha_1$ that of
$\cos\p_a$, then
\BE\label{meanfield}
\cos{\p_a\over 2}\sin{\p_{a+1}\over 2}\sin{\p_{a+2}\over 2}
=\pm\alpha^2\cos{\p_a\over 2}+{\rm fluctuations}
\EE
The sign depend on the relative choice of the ground state for
$\p_{a+1}$ and $\p_{a+2}$. The expectation values $\alpha$ and $\alpha_1$
have been obtained by Lukyanov and Zamolodchikov\cite{Lukyanov97,Lukyanov97B}
and are proportional
to $|m|^{1/4}$ and $m$ respectively. Keeping only the most relevant
terms and neglecting the fluctuations of $\sin\p/2$ and $\cos\p$
around these expectation values, we find the following effective
Lagrangian:
\BEQ\label{lageff}
\Lag_{\rm eff}=\sum_a\Bigl\{&& 
{1\over 8\pi}\l[(\d_\tau\p_a)^2+(\d_x\p_a)^2\r]
-(2m - 16\la_1\alpha_1)\cos\p_a
\pm 4\sqrt 2\la\alpha^2\cos{\p_a\over 2}\NN\\
&&\mp 4\sqrt 2\rho\alpha\cos\t_a\l[\cos{\p_a\over 2}
-\cos{\p_{a+1}\over 2}-\cos{\p_{a+2}\over 2}
\r]\Bigr\}
\EEQ
At this level of approximation, we have three perturbed sine-Gordon models
-- mutually coupled only if $\rho\ne0$ -- and the sign of the interchain
coupling can be incorporated in the choice of ground state. Thus, a
ferromagnetic or antiferromagnetic interchain coupling would have the same
effect. Note that the couplings $\la$ and $\rho$ break the charge conjugation
symmetry (\ref{cconjug}).

\section{Behavior of the gap in the spin ladder}
\label{S-gap}

\begin{figure}
\epsfxsize 8cm\centerline{\epsfbox{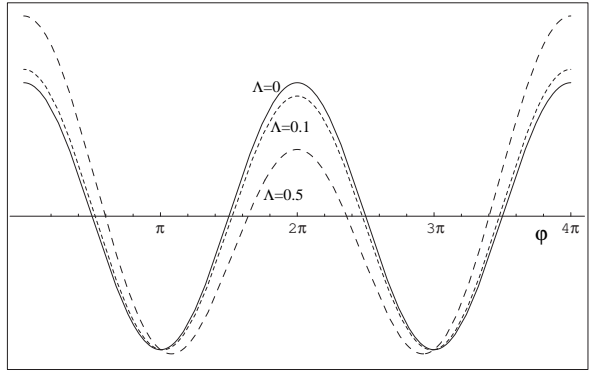}}
\caption{Evolution of the sine-Gordon potential $\cos\p+\La\cos(\p/2)$
for $\La=0$, 0.1 and 0.5. The mass of the
lowest-energy kink decreases linearly with $\La$.}
\label{potFIG}
\end{figure}

Let us first consider the spin-1 ladder, which corresponds to $\rho=0$.
The effective Lagrangian (\ref{lageff}) then reduces to three decoupled,
two-frequency sine-Gordon models:
\BE
\Lag_{\rm lad}=\sum_a\l\{{1\over8\pi}\l[(\d_\tau\p_a)^2+(\d_x\p_a)^2\r]
 -M\cos\p_a
\pm \La\cos{\p_a\over 2}\r\}
\label{laddereff}
\EE
where $M=(2m - 16\la_1\alpha_1)$ and $\La=4\sqrt 2\la\alpha^2$.
This Lagrangian has been studied by Delfino and Mussardo,\cite{Delfino98}
who conclude that, as $\La$ increases from zero, one of the kinks from each
model becomes more massive, whereas the other one becomes less massive. This 
is easily understood by considering the evolution of the potential $V(\p_a)$
as a function of $\La$ (cf Fig.~\ref{potFIG}): the soliton having to bridge
the potential barrier from $\p\sim\pi$ to $\p\sim3\pi$ (towards the right) has
a lower energy than the soliton going from $\p\sim\pi$
to $\p\sim-\pi\equiv3\pi$ (towards the left). Which kink sees its mass 
decrease depends on the sign of the perturbation, but the net result is the
same whatever this sign is. With the help of sine-Gordon form
factors,\cite{Lukyanov97,Lukyanov97B} we can ascertain how the kink mass varies
with $\La$. At first order, the variation of the mass squared
is:\cite{Delfino98}
\BE
\delta m_a^2\approx \vert\La\vert F_{a{\bar a}}(i\pi)
\EE
where the form factor $F$ is
\BE
F_{a{\bar a}}(\eta)\equiv\bra{0}\sin{\p\over 2}~\ket{a(\eta_1){\bar a}
(\eta_2)}
\EE
where $a$ and ${\bar a}$ represent the kink and
antikink and $\eta_{1,2}$ are the associated rapidities
($\eta=\eta_1-\eta_2$). From Ref.~\onlinecite{Lukyanov97}, we extract
the following expression:
\BEQ
F_{a{\bar a}}(\eta)&=-\l\langle \e^{i\p/2}\r\rangle \e^{\eta/2}
-\l\langle \e^{-i\p/2}\r\rangle \e^{-\eta/2}\cr
&=-\l(\hf m_a\r)^{1/4}2^{1/6}A^3\e^{-1/4}2\cosh{\eta\over 2}
\EEQ
where $m_a$ is the mass of the kink and $A\approx1.282427$ is the
Glaisher constant. From this result, we see that $\delta m_a^2$ vanishes at
first order. We thus expect it to be proportional to $\La^2$, This is
compatible with the semiclassical result that the variation of the mass of the
kink is proportional to the variation of the height of the potential. We thus
conclude that
\BE
\delta m_a\propto \La
\EE

According to the analysis of Ref.~\onlinecite{Delfino98}, a single
two-frequency sine-Gordon model has a $c=\frac12$ Ising fixed point, in
addition to the Gaussian fixed point at $M=0$ and $\La=0$.
Since the scaling dimensions of $M$ and
$\La$ at the Gaussian fixed point are respectively
$2-1=1$ and
$2-\frac14=\frac74$, the ratio $\zeta=\La/M^{7/4}$ is invariant under RG
flow and is in fact a control parameter which tells us how far we are from the
Ising fixed point, characterized by a critical value $\zeta_c$. At this
value, i.e., at $\La=\zeta_c M^{7/4}$, the light kinks have exactly zero
mass.  If we return to an Ising-model description of the system, we can
understand intuitively how this flow happens: The effective Lagrangian
(\ref{laddereff}) corresponds to six 2D Ising models coupled pairwise by the
following interaction:
\BE
\Lag_{\rm Ising}= -{\La\over\sqrt2}\sigma\sigma'
\EE
Thus, the excitation such that $\sigma(x)$ is parallel to $\sigma'(x)$
will have a lower mass if $\La>0$ (a similar reasoning holds
when $\La<0$, by changing the sign of $\sigma'$). When $\La$ is large enough,
$\sigma$ must be parallel to $\sigma'$ and this parallel configuration defines
a new Ising model, whose critical point occurs at some value of the ratio
$\La/M^{7/4}$.

\begin{figure}
\epsfxsize 8cm\centerline{\epsfbox{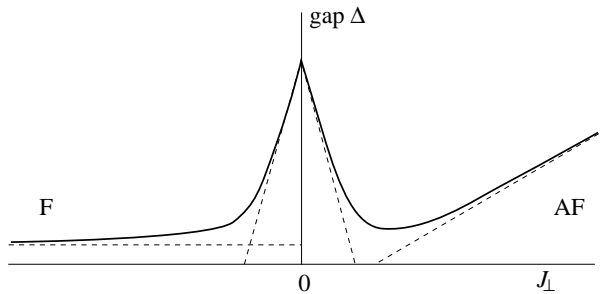}}
\caption{Conjectured dependence of the spin gap $\Delta$ upon the interchain
coupling $\Jperp$ in the spin-1 ladder.}
\label{gapFIG}
\end{figure}
Whereas the mass gap vanishes at a finite value of $\La$ in the effective
Lagrangian (\ref{laddereff}), this does not occur in the spin-1 ladder, because
of the marginal terms $\Lag_2$ and $\Lag_3$ and of the (neglected) fluctuations
of Eq.~(\ref{meanfield}). We nevertheless expect that the spin-1 ladder should
flow close to this fixed point and that the gap should
decrease linearly with a weak interchain coupling, both on the ferromagnetic
and antiferromagnetic sides (with the same slope). This is to be compared with
the Monte-Carlo data of Fig.~3 of Ref.~\onlinecite{Senechal95}, which
illustrates this drop in the gap, for an antiferromagnetic interchain coupling
only. That the gap drops on the ferromagnetic side is not surprising,
considering that (i) the ladder becomes equivalent to a spin-2 chain at large
ferromagnetic coupling and (ii) the gap of an antiferromagnetic
Heisenberg chain with integer spin $s$ decreases with $s$. On the
antiferromagnetic side, the drop in the gap may be roughly understood as a
greater tendency towards magnetic ordering, and should be maximum when
$\Jperp\sim 1$. Beyond $\Jperp\sim 1$, the gap must eventually turn around and
increase linearly at large $\Jperp$, since the lowest-lying excitations are
then rung triplets, costing an energy $\Jperp$. The gap $\Delta(\Jperp)$ is
then conjectured to have a cusp-like maximum at $\Jperp=0$, a peculiar
nonanalytic feature, as illustrated schematically on Fig.~\ref{gapFIG}.

\section{The zigzag spin chain}
\label{S-zigzag}

The zigzag spin-1 chain corresponds to $\delta=0$, and thus $\la=0$,
$\rho\ne0$. The effective Lagrangian is then
\BEQ
\Lag_{\rm zigzag} &=& \sum_a\Biggl\{ {1\over 8\pi}\l[(\d_\tau\p_a)^2
+(\d_x\p)^2\r]-(2m-16\la_1\alpha_1)\cos\p_a\cr
&&\mp 4\sqrt 2\rho\alpha\cos\l(\t_a\r)\l[\cos{\p_a\over 2}
-\cos{\p_{a+1}\over 2}-\cos{\p_{a+2}\over 2}
\r]\Biggr\}
\EEQ
This Lagrangian is not easily analyzed. Let us go back to the
fermionic representation of the twist term by order and disorder fields:
\BE
\Lag_\rho = 2\rho\sum_a \s_a\mu_{a+1}\mu_{a+2}\d_x(\s'_a\mu'_{a+1}\mu'_{a+2})
\EE
With $\L\mu_a\mu'_a\R=-\sqrt2 i\alpha$ (cf Eq.~\ref{bosonising}), the most
relevant term will be
\BE\label{efftwist}
\Lag_\rho \approx -4\alpha^2\rho\,\s_a\d_x\s'_a
\EE
We will now study the effect of this approximate representation of the
twist term by considering the corresponding lattice model (see
Appendix~\ref{appIsing}). Let us map the order fields in the following way:
\BE
\s_a(x)\to \s^x_{a}(n)\qquad
\s'_a(x)\to\s^x_{a}\l(n+{1\over 2}\r)
\EE
With the representation (\ref{efftwist}) for the twist term, the system is
described by the following Hamiltonian:
\BE
H = \sum_{n,a}\l\{-\s^z_a(n/2)
-\kappa\s^x_a(n/2)\s^x_a(n/2+1)\r\}
-4\rho\alpha^2\sum_{n,a}\s^x_a(n)\l[\s^x_a(n+1/2)-\s^x_a(n-1/2)\r]
\EE
where $\kappa$ is related to the constant $m$ (i.e. $-1-\eta$) by the
relation $\kappa= 1+a_0 m$, where $a_0$ is the lattice constant. Thus
$\kappa=1$ for $m=0$ ($\eta=-1$) and tends to 0 when $\eta$ grows.
To bring this Hamiltonian to a more familiar form, we perform a rotation
of $\pi$ around the $z$ axis of the spin operator at every other site, on each
chain. This changes the sign of $\s^x_a$ on those sites and gives the
Hamiltonian a slightly different form:
\BE \label{effzigzag}
H = \sum_{n,a}\l\{-\s^z_a(n/2)
+\kappa\s^x_a(n/2)\s^x_a(n/2+1)\r\}
+4\rho\alpha^2\sum_{n,a}\s^x_a(n)\l[\s^x_a(n+1/2)+\s^x_a(n-1/2)\r]
\EE

\begin{figure}
\epsfxsize 8cm\centerline{\epsfbox{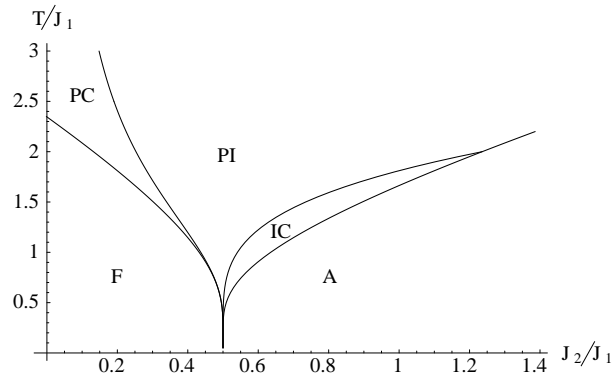}}
\caption{Phase diagram of the classical ANNNI model. $T$ is the temperature,
$J_1$ and $J_2$ are the nearest-neighbor (NN) and next-nearest-neighbor (NNN)
Ising couplings, respectively.}
\label{phasani}
\end{figure}

The Hamiltonian (\ref{effzigzag}) defines the quantum ANNNI model. Together
with its two-dimensional, classical counterpart (cf
Refs~\onlinecite{Villain81,Rujan81,Selke88}), it has been extensively studied
by a variety of methods: mean field theory,\cite{vonBoehm79} Monte Carlo
simulations,\cite{Hornreich79,Selke80,Selke81,Arizmendi91}
Muller-Hartmann-Zittartz approximation,\cite{Hornreich79} perturbative
expansions,\cite{Barber81,Barber82,Barber82B} free fermion
approximation\cite{Villain81,Rujan81,Rieger96} and exact
diagonalizations.\cite{Rieger96,Sen95} The phase diagram for the classical
model is shown in Fig.~\ref{phasani}. In the scaling limit,
the temperature $T$ of the classical model model is related to the mass $m$ by
$T=(1-a_0 m)T_c=(2-\kappa)T_c$. The NN coupling is proportionnal to the
interchain coupling $\rho$ and $J_2$ to $\kappa$. Thus, the case of small
zigzag interaction corresponds to the limit of small $J_1$. The different
phases are the following: ferromagnetic (F), paramagnetic commensurate (PC),
paramagnetic incommensurate (PI), incommensurate critical phase (IC, also
called ``floating phase'') and antiphase (A) of alternating pairs
($++--++\cdots$). A disorder line found by Peschel and Emery\cite{Peschel81}
divides the PC and the PI phase.

We conclude from this phase diagram that incommensurability will arise in the
spin-1 zigzag chain as soon as the interchain coupling is nonzero
(the model (\ref{effzigzag}) is then in the far-right of the PI phase). One
premise for this deduction is that the incommensurability of the Ising spins
($\sigma_a$) is reflected in the correlation of the spins of the quantum
chain; this comes from the relation (\ref{WZWisinge}). Note that increasing
$\rho$ brings us from infinity on the phase diagram~\ref{phasani} towards the
origin, along a straight line. One could expect such a line to go through 
other phases (like the IC phase) at some point. However, we should note that
the omission of the fluctuation of $\mu_a$ in Eq.~(\ref{efftwist}) is valid
only when $T-T_c$ is large compare to $\rho$.

Moreover, we can have an idea of how the incommensurability develops as a
function of $\rho$. A recent analysis using a high-temperature expansion and
bosonisation\cite{Allen99B} shows that in the limit of very strong
next-nearest-neighbor interaction in the ANNNI model, the incommensurability
is proportional to $\rho/\kappa$. Explicitely, in the high temperature
limit, the incommensurate wavevector is given by:
\BE
q_0=\pm{2\alpha^2\rho(1+\kappa+\kappa^2+\cdots)\over 2\kappa(1-\kappa)}
\EE
where the dots stand for higher power of $\kappa$. The $\pm$ sign are
respectively associated to the correlation function of the combination
$\pm\s^z_{a}(n)+\s^z_{a}(n\mp 1/2)$. This result that the
incommensurability is linear with the interchain coupling confirms the
one obtained by a semiclassical analysis.\cite{Allen95}

\acknowledgments

D.A. thanks Philippe Lecheminant for useful discussion. This work was
partially supported by NSERC (Canada) and by le Fonds FCAR (Qu\'ebec).

\appendix
\section{The 2D Ising model}
\label{appIsing}

In this appendix, we review briefly the correspondence of the Ising model
with fermions, the conformal structure of the model and we indicate a
set of careful bosonization formulas for a pair of Ising models.

\subsection{Definitions}

As is well known, the 2D statistical Ising model is equivalent to a
quantum Ising chain in a transverse field, with
Hamiltonian
\BE\label{isingtrans}
H=-\la\sum_i\s^z_i-\sum_i\s^x_{i}\s^x_{i+1}
\EE
where $\s^{1,2,3}$ are the Pauli matrices.  The Hamiltonian (\ref{isingtrans})
can be diagonalized through a Jordan-Wigner transformation followed by a
Bogolubov-Valatin transformation. The solution shows that $\L\s_i^x\R\ne0$ if
$\la<1$ and $\L\s_i^x\R=0$ otherwise. Thus $\la=1$ is the critical point.
A peculiarity of this model is the existence of a duality transformation
mapping the ordered phase to the disordered phase and vice-versa.
Under this transformation the spin operators $\s_i^a$ are mapped to the
so-called disorder operators, defined on links (dual lattice) by the
following relations:
\BEQ\label{disord}
\mu_{i+1/2}^z &=&\s^x_{i}\s^x_{i+1}\NN\\
\mu_{i+1/2}^x &=&\prod_{j\leq i}\s_j^z
\EEQ
Let us apply the Jordan-Wigner transformation on the dual lattice. The fermion
creation and annihilation operators are defined as:
\BEQ\label{JordWigner}
c_{j+1/2} &=& \mu^{-}_{j+1/2}\exp\l( 
{i\pi\over2}\sum_{k<j}(\mu_{k+1/2}^z-1) \r)\NN\\ 
c^\dagger_{j+1/2} &=& \mu^{+}_{j+1/2}\exp\l(
{-i\pi\over 2}\sum_{k<j} (\mu_{k+1/2}^z-1) \r)
\EEQ
where $\mu^{\pm}=(\mu^x\pm i\mu^y)/2$. The fermions
correspond to the kinks in the original formulation. Indeed the
fermion number on link $j+1/2$ is
\BE
c_{j+1/2}^\dagger c_{j+1/2} = {1-\s^x_{j}\s^x_{j+1}\over 2}~,
\EE
i.e., there is no fermion on the link if the spins on $j$ and
$j+1$ are parallel and one if they are antiparallel. Note that the order
parameter $\s^x$ has a bosonic character, whereas the disorder parameter
$\mu^x$ is fermionic. This is easily seen from the following equivalence:
\BEQ
\mu^x_{j+1/2} &=& c^\dagger_{j+1/2}\exp\l(
-i\pi\sum_{k<j}c^\dagger_{k+1/2}c_{k+1/2}\r)
+c_{j+1/2}\exp\l(i\pi\sum_{k<j}c^\dagger_{k+1/2}c_{k+1/2}\r)\NN\\
\s^x_j &=& \s^x_{-N}\exp\l(\pm i\pi\sum_{k<j}c^\dagger_{k+1/2}c_{k+1/2}\r)
\EEQ

\subsection{Continuum limit}

The critical point of the 2D Ising model is equivalent, in the continuum limit,
to a free, massless Majorana fermion: a conformal field theory with central
charge $c=1/2$ and three conformal families: the identity operator,
the energy operator $\epsilon$, and the the spin density (or order) operator
$\s$. The use of complex coordinates $z=\tau+ix$ and $\zb=\tau-ix$
is standard, along with the complex derivatives $\d=\d_z=(\d_\tau-i\d_x)/2$
and $\bar\d=\d_\zb=(\d_\tau+i\d_x)/2$ (the notation used is that of
Ref.~\onlinecite{DMS97}). The energy density operator may be expressed in
terms of the chiral components of the Majorana fermion as $\epsilon =
i\psi\psib$. The order field $\s$ is the continuum limit of the spin operator
$\s_i^x$, and a fermionic disorder field $\mu$ may be introduced as the
continuum limit of the disorder operator $\mu_{i+1/2}^x$. The field $\mu$ has
the same scaling properties as the field $\s$, but is nonlocal with respect to
$\s$. The conformal transformations are generated by the energy-momentum tensor, whose
chiral components are $T =-\hf\psi\d\psi$ and $\bar T =-\hf\psib\bar\d\psib$.
All these fields have the following short-distance products, or operator
product expansion (OPE):
\BEQ\label{OPEising}
\psi(z)\psi(w) &\sim& {1\over z-w}+2(z-w)T(w) \NN\\
\psib(\zb)\psib(\wb) &\sim& {1\over\zb-\wb}+2(\zb-\wb) {\bar T}(\wb) \NN\\
\s(z,\zb)\s(w,\wb) &\sim& {1\over\vert z-w\vert^{1/4}}+{1\over 2}
\vert z-w\vert^{3/4}\epsilon(w,\wb)\NN\\
\mu(z,\zb)\mu(w,\wb) &\sim& {1\over\vert z-w\vert^{1/4}}-{1\over 2}
\vert z-w\vert^{3/4}\epsilon(w,\wb)\NN\\
\s(z,\zb)\mu(w,\wb) &\sim& {\gamma(z-w)^{1/2}\psi(w)
+\gamma^* (\zb-\wb)^{1/2}\psib(\wb)\over\sqrt 2\vert
z-w\vert^{1/4}}\NN\\
\mu(z,\zb)\s(w,\wb) &\sim& {\gamma^*(z-w)^{1/2}\psi(w)
+\gamma (\zb-\wb)^{1/2}\psib(\wb)\over\sqrt 2\vert z-w\vert^{1/4}}\NN\\
\psi(z)\s(w,\wb) &\sim& {\gamma\over\sqrt 2(z-w)^{1/2}}\mu(w,\wb)\NN\\
\psi(z)\mu(w,\wb) &\sim& {\gamma^*\over\sqrt
2(z-w)^{1/2}}\s(w,\wb)\NN\\
\psib(\zb)\s(w,\wb) &\sim& {\gamma^*\over\sqrt 2(\zb-\wb)^{1/2}}
\mu(w,\wb)\NN\\
\psib(\zb)\mu(w,\wb) &\sim& {\gamma\over\sqrt 2(\zb-\wb)^{1/2}}
\s(w,\wb)
\EEQ
where $\gamma=\exp{i\pi/4}$ (or, equivalently, $\exp{-i\pi/4}$).

\subsection{Bosonization}

Two Ising models form a $c=1/2+1/2=1$ conformal theory. We therefore expect a
representation of the different fields in terms of a free boson $\p$, defined
by the action 
\BE
S={1\over 8\pi}\int \dr x\dr\tau\l[ \l(\d_\tau\p\r)^2 +\l(\d_x\p\r)^2\r]
\EE
Our choice of normalization ($1/8\pi$) simplifies the exponentials and
circular functions appearing in the sine-Gordon theory. The OPE of the
boson field and of its (normal-ordered) exponentials are
\BEQ
\d\p(z)\d\p(w) &\sim& {1\over(z-w)^2}\NN\\
\e^{i\alpha\p(z)}\e^{i\beta\p(w)}
 &\sim& (z-w)^{\alpha\beta}\e^{i(\alpha+\beta)\p(w)}+\cdots
\EEQ
The boson field can be separated into chiral components:
$\p(x,\tau) = \phi(z)+\phib(\zb)$. These fields have the following
mode expansion in radial quantization:\cite{DMS97}
\BEQ
\phi(z) &=&q-ip\log z +i\sum_{k\neq 0}{1\over k}a_k z^{-k} \NN\\
\phib(\zb)&=&\qb-i\pb\log\zb +i\sum_{k\neq 0}{1\over k}\ab_k \zb^{-k}
\EEQ
where the operators $p$, $q$, $a_n$ satisfy the following commutation
relations:
\BE
[q,p]=i\qquad [a_n,a_m]=n\delta_{n,m}
\EE
with similar relations for the left-moving (barred) operators. 

A faithful representation of the Ising fields is then given by the following
relations:\cite{Allen98}
\BEQ
\label{bosonising}
\psi&=&{1\over\sqrt 2}\e^{i\pi\pb}\e^{i\phi}+ {1\over\sqrt
2}\e^{-i\pi\pb}\e^{-i\phi}\NN\\
\psib&=&-{i\over\sqrt 2}\e^{-i\pi\pb}\e^{-i\phib}- {i\over\sqrt
2}\e^{i\pi\pb}\e^{i\phib}\NN\\
\psi'&=&-{i\over\sqrt 2}\e^{i\pi\pb}\e^{i\phi}+ {i\over\sqrt
2}\e^{-i\pi\pb}\e^{-i\phi}\NN\\
\psib{}'&=&-{1\over\sqrt 2}\e^{-i\pi\pb}\e^{-i\phib}+ {1\over\sqrt
2}\e^{i\pi\pb}\e^{i\phib}\NN\\
\s\s'&=&\sqrt 2\cos{\p\over 2}\NN\\
\s\mu'&=&-i\sqrt 2\sin{\theta\over 2}\NN\\
\mu\s'&=&-i\sqrt 2\cos{\theta\over 2}\NN\\
\mu\mu'&=&-\sqrt 2 i\sin{\p\over 2}
\EEQ
where $\t=\phi-\phib+2\pi\pb$ is the field dual to $\p$ (the operator $2\pi\pb$
is added to ensure proper anticommuation properties). This representation leads
to the correct OPE (\ref{OPEising}) between the Ising fields.
The phase factor $\e^{i\pi\pb}$ is similar to the phase factor in the
Jordan-Wigner transformation, $\pb$ being to the number of left fermions.
Only its odd or even character matters. We note the natural periodicity
property $\p\to\p+4\pi$ and $\t\to\t+4\pi$ of this representation.

The energy-momentum tensors $T$ and $T'$ of the two Ising models, along
with their antiholomorphic counterparts, are bosonized as follows:
\BEQ\label{EMtens}
T(z)-T'(z) &=& 4\sqrt 2 \e^{2\pi i\pb}\cos\l( 2\phi(z)\r)\NN\\
{\bar T}(\zb)-{\bar T}'(\zb) &=&
-4\sqrt 2 \e^{2\pi i\pb}\cos\l( 2\phib(\zb)\r)
\EEQ
This relation is useful when bosonizing the twist term (the last term of
Eq.~(\ref{lagcoupled})).


\end{document}